# Topological states

Dimitrie Culcer[1, 2]
[1]School of Physics, The University of New South Wales, Sydney 2052, Australia
[2]Australian Research Council Centre of Excellence in Low-Energy Electronics Technologies, UNSW Node, The University of New South Wales, Sydney 2052, Australia

Attila Geresdi[3]
[3]QuTech and Kavli Institute of Nanoscience, Delft University of Technology, 2600 GA Delft, The Netherlands

**Status**
Topological phases are characterised by a topological invariant that remains unchanged by deformations in the Hamiltonian. Materials exhibiting topological phases include topological insulators (TI), superconductors exhibiting strong spin-orbit coupling, transition metal dichalcogenides, which can be made atomically thin and have direct band gaps, as well as high mobility Weyl and Dirac semimetals. In 2D topological materials, the electron gas on the surface has enabled spectacular phenomena such as the spin-orbit torque: a current through a TI can flip the magnetisation of an adjacent ferromagnet even at room temperature [1]. A power-saving topological transistor harnesses edge states that conduct electricity without dissipation and are responsible for the observed quantised spin and anomalous Hall effects. The first topological transistor design exploits the fact that a top gate drives the system between an insulating normal phase and a topological phase with a quantised conductance [2]. Non-linear electrical and optical effects have taken off, with grand aims including identifying a Hall effect in time-reversal symmetric systems [3] and a direct photocurrent, the *shift current*, which could enable efficient solar cell paradigms.

Topological quantum computing is another motivation to investigate this field. The current physical realisations of quantum bits invariably suffer from finite control fidelity and decoherence due to interaction with the environment. While error correction schemes have been developed to tackle these challenges, topologically protected states can mitigate it by harnessing their degenerate ground state manifold. Topologically protected quantum operations on this manifold require quasiparticles with non-Abelian exchange statistics, which emerge in various engineered nanostructures where electrons are confined to one or two dimensions. Importantly, it is this reduced dimensional behaviour that enables non-Abelian exchange statistics [4]. Specifically, planar semiconductor heterostructures hosting the $\nu = 5/2$ fractional quantum Hall state [4] and spin-orbit coupled nanowires with induced superconductivity in an external magnetic field [5] are widely investigated platforms (Fig. 1). In these devices, quasiparticle tunnelling [6] and Josephson effect experiments [7] resulted in signatures consistent with the presence of topologically protected electron states. However, no unambiguous experimental confirmation of the non-Abelian exchange statistics exists to date. This next step, which is a prerequisite of topological quantum computation, relies on progress in materials science and the development of readout schemes.

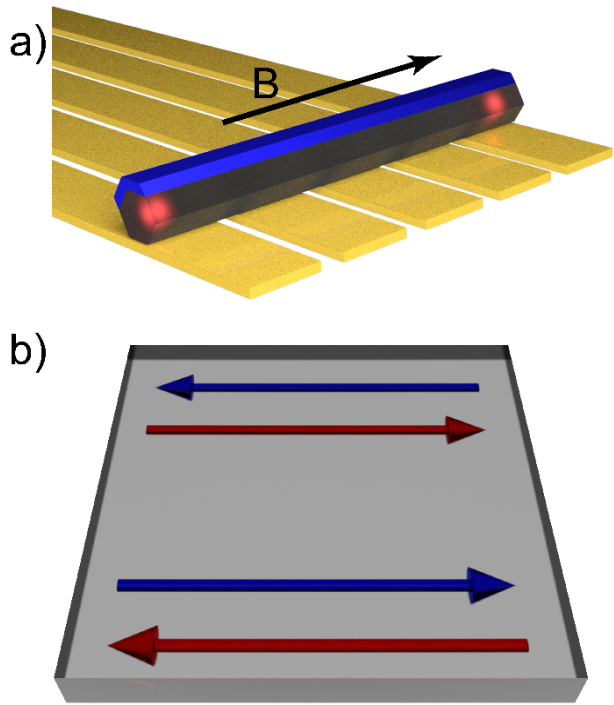

**Figure 1.** Fundamentals of topologically protected electron states. (a) Majorana zero modes (in red) are localised at the ends of a semiconductor nanowire (in grey) with superconductivity induced by thin superconducting layers (in blue). The chemical potential can be adjusted by local electrostatic gates (golden stripes), and an external magnetic field B is applied [5]. (b) The quantum spin Hall (QSH) state, where spin-momentum locking of the edge modes (red and blue arrows) prevents backscattering in the absence of time-reversal breaking perturbations. Note that in the case of the quantum anomalous Hall effect (QAHE), there is only one spin-polarised edge mode (either blue or red) and there is no backscattering whatsoever.

**Current and Future Challenges**

Because topological transistor research focuses on the dissipationless transport at low temperatures, while spin-orbit torque devices aim for room-temperature operation, significant gaps exist in our overall understanding of topological material devices. The origin of the experimentally observed strong spin-orbit torque, whether from electrically generated spin polarisations on the surface or spin currents in the bulk, is unclear. For topological transistors, the challenge is to make the threshold voltage as small as possible, below that of conventional transistors. Magnetic ordering, critical for certain topological transistors and for room-temperature operation, is also poorly understood in topological materials. For example, known TIs are layered materials, and it is unknown whether intra-layer or inter-layer magnetic interactions dominate, while different mechanisms may be responsible for magnetic ordering on the surface and in the bulk, resulting in different critical temperatures. Moreover, whether the edge states of TIs are topologically protected, remains controversial. Backscattering of edge states, which destroys the conductance quantisation, can be induced by impurities, either directly or by inducing coupling to bulk states, or by size quantisation effects: the edge itself may experience spontaneous time-reversal symmetry breaking due to edge reconstruction. In extreme cases this leads to Anderson localisation of the edge states. The surface states themselves are sensitive to the metallic contacts on the TI.

Currently, topologically protected electron states are probed by low frequency electronic transport, and high frequency techniques, such as charge sensing, shot noise experiments and the AC Josephson effect. All of these experimental techniques yield signatures of the topologically phase transition, however these signatures were theoretically shown to persist without any topologically protected edge state as well [8]. Future measurements should thus focus on the non-local nature of the topological ordering instead of the local mapping of the edge modes.

A common requirement for topologically protected quantum electronics is the coexistence of a bulk gapped state with the protected edge modes. This requirement demands a careful engineering of materials. Steps in direction have already been made, such as superconductor gap engineering, intentional doping [9] and atomically clean heterointerfaces.

**Advances in Science and Technology to Meet Challenges**
While promising for applications, harnessing topological protection with a technological relevance requires further progress to address the limitations and challenges discussed above. Reducing the threshold voltage of a topological transistor is expected to be a matter of identifying the optimal materials, while recent work suggests that the operating temperature of topological transistors can be enhanced by compensated *n – p* co-doping [9]. Further experiments are needed to resolve the issue of topological protection, since not enough devices exist to determine whether it is fundamental or not, and theoretical predictions of Anderson localisation have not been confirmed or denied. Understanding charge and spin dynamics in the vicinity of interfaces between TIs and ferromagnets is vital in interpreting experiments. Whereas these can be simulated using state-of-the-art computational approaches, they also require conceptual advances in transport theory, considering fundamental issues such as the definition of the spin current when the spin is not conserved. In materials science, required progress includes the growth of clean materials with tailored band parameters to optimize the topological energy gap separating the ground state manifold and trivial excited states. Heterointerface engineering will enable tuning of proximity effects of superconductivity, magnetic ordering or spin-orbit coupling at interfaces, leading to true designer material systems. These developments, which involve complex charge redistributions and modifications of the electronic structure, will require synergy with advanced numerical modelling methods, where progress is limited due to the numerically expensive calculation of geometries including heterostructures.

While topological device schemes have already been suggested (Fig. 2), further investigations are required to address connectivity issues in scalable systems. Specifically for topological quantum bit schemes, existing proposals often neglect electrostatic gating and auxiliary structures required for state preparation and readout.

Finally, advances in quantum algorithms are required to consider specific topological quantum hardware and has to optimize the usage of topologically not protected quantum gates, such as non-Clifford gates for Ising anyons [4]. On the other hand, if experimentally observed, more complex topological states, such as Fibonacci anyons can circumvent this requirement, as they enable universal and topologically protected quantum computation [4].

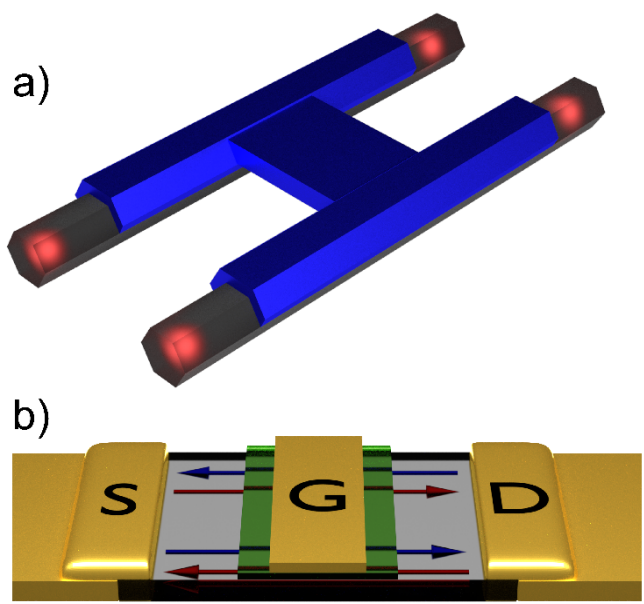

**Figure 2.** Topological device concepts. (a) The Majorana box qubit [10], where the quantum information is encoded in the joint parity of four Majorana states (red spheres). The nanowires are connected via a superconducting bridge (in blue) in order to form a single superconducting island, yet prevent parity leakage. (b) The topological spin transistor, which relies on the gate-tunability of the topological phase transition, recently demonstrated in [2].

## Concluding Remarks

Topological states form an active field with tremendous promise in fundamental science and conventional and topological quantum computing, which will continue to grow. In transport devices, where attention has focused on ferromagnets, antiferromagnets are evolving at a fast pace, and skyrmions offer new avenues for exploiting topological effects. Spin-orbit coupling offers new functionalities for magneto-resistive devices, such as spin valves, since electrons travelling in a specific direction have a fixed spin orientation determined by their momentum. In non-linear response much work remains to be done in understanding the interplay of topological mechanisms with disorder and phonons, and developing a unified theoretical description in terms of doping, disorder correlations and inter-band coherence. Topologically protected quantum information processing, once demonstrated, will have a major impact on scalable quantum technologies, and will provide new directions for the development of quantum algorithms. Finally, there is a lot of space to investigate layered van der Waals heterostructures where experiments are just beginning.

## Acknowledgements

DC is supported by the Australian Research Council Centre of Excellence in Future Low-Energy Electronics Technologies (project CE170100039) funded by the Australian Government. AG acknowledges funding by the European Research Council under the European Union's Horizon 2020 research and innovation programme, grant no. 804988 (SiMS).

## References

[1]     Y. Wang, Nature Comm. **8**, 1364 (2017); J. Han et al, Phys. Rev. Lett. **119**, 077702 (2017)
[2]     J. L. Collins et al, Nature **564**, 390 (2018)
[3]     Q. Ma et al, Nature **565**, 337 (2019); K. Kang, et al, Nature Mater. **18**, 324 (2019)

[4]     C. Nayak et al, Rev. Mod. Phys. **80**, 1083 (2008)
[5]     R. Lutchyn et al, Phys. Rev. Lett. **105**, 077001 (2010); Y. Oreg et al, Phys. Rev. Lett. **105**, 177002 (2010)
[6]     V. Mourik et al, Science **336**, 1003 (2012)
[7]     D. Laroche et al, Nature Comm. **10**, 245 (2019)
[8]     A. Vuik et al, arXiv:1806.02801 (2018)
[9]     S. Qi, et al, Phys. Rev. Lett. **117**, 056804 (2016)
[10]    S. Plugge et al, New Journal of Physics **19**, 012001 (2017)